\documentclass[runningheads]{llncs}
\usepackage{graphicx}
\usepackage{xcolor}
\usepackage[multi-part-units=single]{siunitx}
\usepackage{amsmath}
\usepackage{amssymb}
\usepackage{bm}
\begin{document}
\title{Spectral-Spatial Recurrent-Convolutional Networks for \textit{In-Vivo} Hyperspectral Tumor Type Classification}
\titlerunning{\textit{In-vivo} hyperspectral spectral-spatial deep learning}
%
\author{Marcel Bengs\inst{1}$^*$ \and Nils Gessert\inst{1}$^*$ \and Wiebke Laffers\inst{2}\and Dennis Eggert\inst{3} \and Stephan Westermann\inst{4} \and
Nina A. Mueller\inst{4} \and Andreas O. H. Gerstner\inst{5} \and Christian Betz\inst{3} \and
Alexander Schlaefer\inst{1}}

\index{Bengs, Marcel}
\index{Gessert, Nils}
\index{Laffers, Wiebke}
\index{Eggert, Dennis}
\index{Westermann, Stephan}
\index{Mueller, Nina A.}
\index{Gerstner, Andreas O. H.}
\index{Betz, Christian}
\index{Schlaefer, Alexander}
\authorrunning{Bengs et al.}
%
\institute{Institute of Medical Technology and Intelligent Systems, Hamburg University of Technology, Germany \and
Departments of Otorhinolaryngology/Head and Neck Surgery, Carl von Ossietzky University Oldenburg, Germany \and
Clinic and Polyclinic for Otolaryngology, University Medical Center Hamburg-Eppendorf, Germany \and 
Department of Otorhinolaryngology/Head and Neck Surgery, University of Bonn, Germany \and ENT-Clinic, Klinikum Braunschweig, Germany\\
$^*$ Both authors contributed equally.\\ 
\email{marcel.bengs@tuhh.de}}
\maketitle              
\begin{abstract} 
Early detection of cancerous tissue is crucial for long-term patient survival. In the head and neck region, a typical diagnostic procedure is an endoscopic intervention where a medical expert manually assesses tissue using RGB camera images. While healthy and tumor regions are generally easier to distinguish, differentiating benign and malignant tumors is very challenging. This requires an invasive biopsy, followed by histological evaluation for diagnosis. Also, during tumor resection, tumor margins need to be verified by histological analysis. To avoid unnecessary tissue resection, a non-invasive, image-based diagnostic tool would be very valuable. Recently, hyperspectral imaging paired with deep learning has been proposed for this task, demonstrating promising results on \textit{ex-vivo} specimens. In this work, we demonstrate the feasibility of \textit{in-vivo} tumor type classification using hyperspectral imaging and deep learning. We analyze the value of using multiple hyperspectral bands compared to conventional RGB images and we study several machine learning models' ability to make use of the additional spectral information. Based on our insights, we address spectral and spatial processing using recurrent-convolutional models for effective spectral aggregating and spatial feature learning. Our best model achieves an AUC of $\SI{76.3}{\percent}$, significantly outperforming previous conventional and deep learning methods.

\end{abstract}
%
%
\section{Introduction}
Head and neck cancers are responsible for 3.6$\%$ of cancer-specific deaths while being the sixth most common type of cancer \cite{shield2017global}. For a patient's prognosis, early detection is critical \cite{horowitz2001perform}. Late detection of malignancy has been reported for tumors in the head and neck area \cite{habermann2001carcinoma}, suggesting that accurate diagnostic tools would be valuable. The typical diagnostic procedure for head and neck cancer starts with an endoscopic intervention where a medical expert examines tissue regions. In case malignancy is suspected for a tumor region, an invasive biopsy, followed by histological evaluation is performed. While being considered the gold standard for diagnosis, tissue removal can cause function deterioration \cite{alieva2018potential}. Also, a biopsy is time-consuming and a costly procedure that needs to be performed under anesthesia, leading to a risk for the patient. 

As a consequence, fast and accurate, non-invasive diagnosis or patient referral would have the potential to significantly improve the head and neck cancer diagnostic procedure both for patients and medical staff. For this purpose, several optical imaging techniques, embedded into an endoscope, have been studied over the recent years. For example, optical microscopy \cite{lohler2014incidence}, optical coherence tomography using infrared light \cite{arens2007indirect} and narrow-band imaging augmented by fluorescence \cite{volgger2013evaluation} have been studied for detection of malignancy.

Another promising imaging modality is hyperspectral imaging (HSI) where multiple images are acquired at several different wavelengths of light. The method has also been studied in the context of head and neck cancer detection \cite{regeling2016hyperspectral,regeling2016development}. This includes an \textit{in-vivo} study where conventional machine learning methods were employed for pixel-wise tissue classification using a single patient for training \cite{laffers2016early}. Very recently, deep learning methods have been employed for HSI-based head and neck cancer detection in an \textit{ex-vivo} setting \cite{halicek2018optical,grigoroiu2020deep}. Also, distinguishing healthy tissue from tumor areas has been studied with different convolutional neural networks (CNNs) for spatio-spectral HSI processing \cite{bengs-2020-1,eggert2020vivo}. In this work, we study the more challenging task of \textit{in-vivo} tumor type classification using HSI and different machine learning methods. 

In terms of methods, a straight-forward solution is stacking all spectral bands in the CNN's input channel dimension and assuming the spectral bands to be similar to RGB color channels. However, several studies have demonstrated that joint spatio-spectral deep learning can substantially improve classification performance. For example, Liu et al. used convolutional long short-term memory (LSTM) models instead of CNNs, adopting recurrent processing for temporal sequences to spectral bands \cite{liu2017bidirectional}. Also, a 3D extension of the popular Inception architecture for joint spatial and spectral processing where all dimensions are treated with convolutions has been employed \cite{halicek2018optical}. However, there have hardly been any studies detailing the individual importance of spectral and spatial information. Also, it is often unclear how much additional value multiple spectral bands provide over RGB images and whether machine learning methods are able to use the additional spectral information effectively.

Therefore, we compare both conventional machine learning approaches and deep learning methods using only spectral, mostly spatial or combined spatial and spectral information. To maximize use of all available information, 3D CNNs with convolution-based processing for both spatial and spectral dimensions should be able to extract rich features. However, the approach assumes similar importance of spectral and spatial information although there is a high correlation and similarity between spectral bands. We hypothesize that learning a compact spectral representation first, followed by a focus on spatial processing is more effective. Therefore, we propose the use of recurrent-convolutional models with a convolutional gated recurrent unit (CGRU) followed by a CNN. The CGRU first aggregates the 3D hyperspectral cube into a 2D representation which is then processed by a 2D CNN. We provide an analysis of several architectural variations of the proposed model and show how changes in the number of spectral bands affect performance.

Summarized, our contributions are three-fold. First, we demonstrate the feasibility of deep learning-based \textit{in-vivo} head and neck tumor type classification in hyperspectral images. Second, we provide an extensive analysis of different models for evaluating the spectral and spatial information content. Third, we propose a convolutional-recurrent model for efficient spectral and spatial processing. 

\section{Methods}

\subsection{Data Set}

\begin{figure}[t]
\centering
\includegraphics[width=0.34\textwidth]{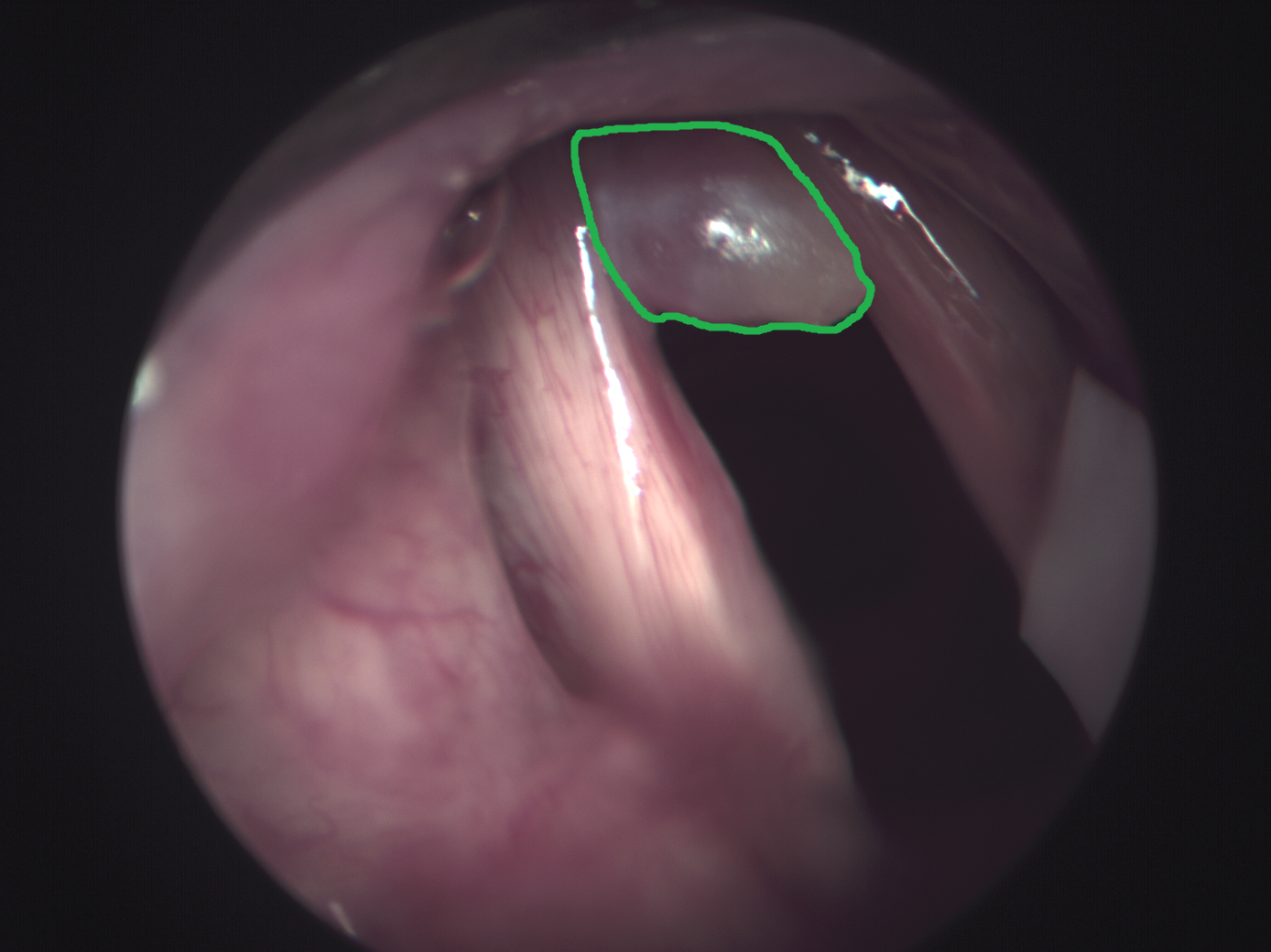}
\includegraphics[width=0.34\textwidth]{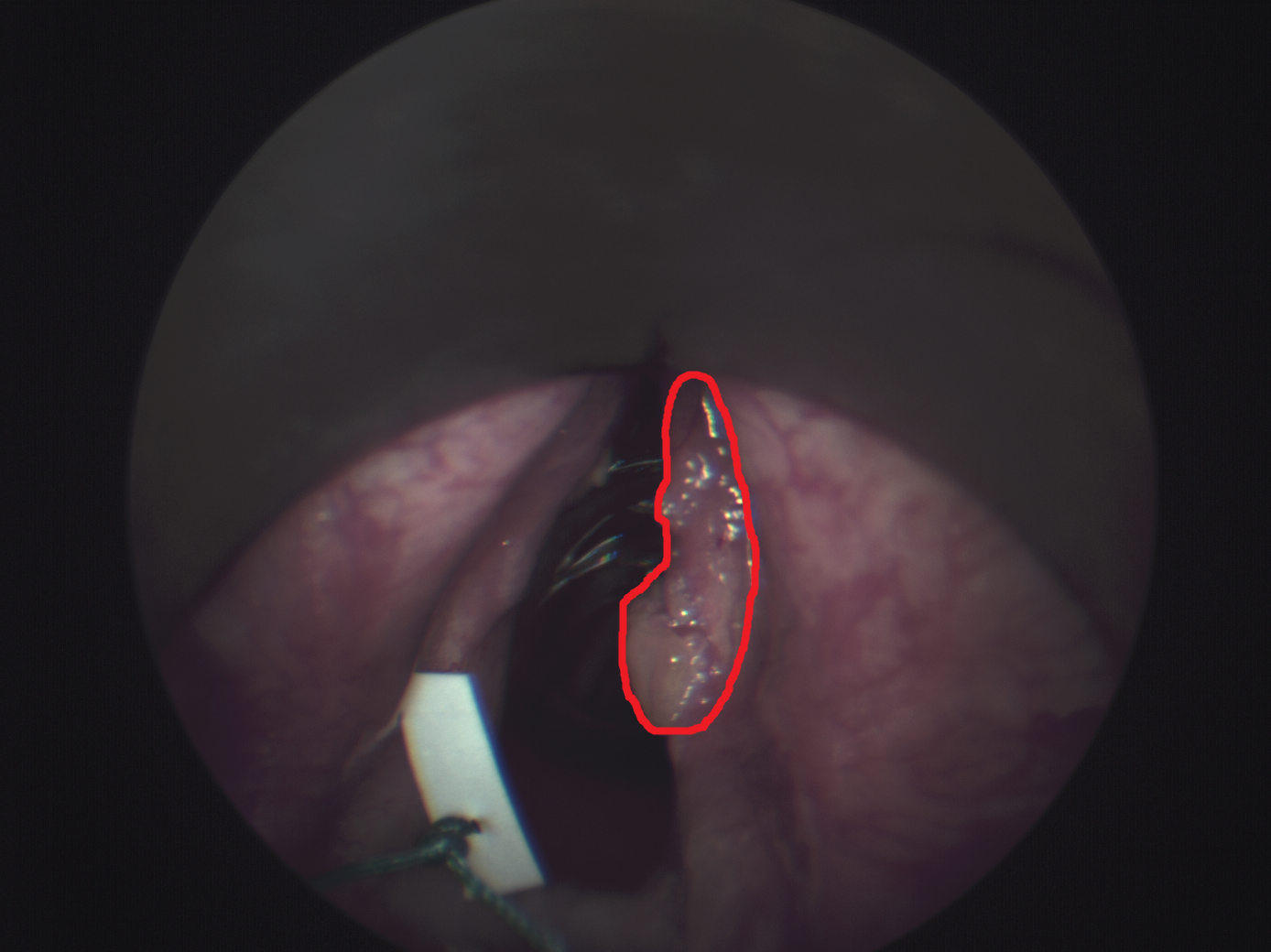}
\caption{Example RGB images including marked tumors. Left, a bengin tumor at the vocal folds is shown, right, a malignant tumor at the vocal folds is shown.}
\label{fig:examples}
\end{figure}
The data set consists of 98 patients, who were examined in a previous study due to mucous membrane abnormalities in the area of the upper aerodigestive tract \cite{bengs-2020-1}. Data acquisition was performed as described by Gerstner et al. \cite{gerstner2012hyperspectral} at the department of otorhinolaryngology at the University of Bonn. The study was approved by the local ethics committee (176/10 \& 061/13, University of Bonn). The acquisition was performed throughout a normal procedure for diagnosis. An endoscopic device (Karl Storz GmbH \& CoKG, Tuttlingen, Germany) was used for visual assessment. A Polychrome V monochromator (TillPhotonics, Gr\"afelfing, Germany) was used as light source for HSI and a monochromatic CCD-camera (AxioCamMRm, Carl Zeiss Microimaging GmbH, G\"ottingen, Germany) was employed for HSI data acquisition. The spectral bands range from $\SI{430}{\nano\metre}$ to $\SI{680}{\nano\metre}$ with a step size of $\SI{10}{\nano\metre}$ and a bandwidth of $\SI{15}{\nano\metre}$. The spatial resolution is $1040\times1388$ pixels. Biopsy samples were extracted from the imaged tumor areas, followed by histopathological evaluation. After the acquisition, we use the ImageJ-implementation of the SIFT-algorithm “Linear Stack Alignment with SIFT” \cite{lowe2004distinctive} for aligning of 
the HSI images. Experts performed ground-truth annotation using RGB representations derived from the HSI images and the histopathological report. Note that this method induces some label noise as tumor outlines might be slightly inaccurate. In total, there are 83 patients with benign tumor region and 15 patients with malignant tumor region. Thus, the learning problem is particularly challenging due to extreme class imbalance. Tissue regions include the larynx, oropharynx and hypopharynx. Example images and highlighted tumor regions are shown in Figure~\ref{fig:examples}. For model training and evaluation, we crop patches of size $32\times 32$ from the marked areas, including a margin towards the marked border. Overall, we obtain $\num{18025}$ patches from all patients. For training and evaluation of our models, we define three subsets with a size of 19 patients (5 malignant / 14 benign) each and apply a cross-fold scheme. Note, data from one patient does not appear in different subsets. We split each subset into a test set (3 malignant / 8 benign) and validation set (2 malignant / 6 benign). We perform hyperparameter optimization using grid search on the validation splits and we report performance metrics for the test splits.

\subsection{Machine Learning and Deep Learning Methods}

We consider several different machine learning methods for \textit{in-vivo} head and neck tumor type classification. A first baseline is the use of a random forests (\textbf{RF}) and support vector machines (\textbf{SVM}), similar to a previous approach where a single patient was used for training \cite{laffers2016early}. We use the spectral bands as features for the RF and SVM and perform pixel-wise classification. Thus, these conventional models only use spectral information and no spatial context. In contrast, deep learning methods such as CNNs are designed for spatial processing. A first straightforward approach is to use a 2D CNN with RGB color images, which can be derived from the HSI stacks. Here, all color channels are stacked into the first layer's channel dimension which is the standard approach for natural color images. We refer to this approach as \textbf{2D CNN RGB}. This method serves as a baseline for the use of mostly spatial information and no additional spectral bands. The approach can be directly extended to the use of hyperspectral images by treating the additional spectral bands as colors channels. Thus, here, we also stack all $N_S$ spectral bands into the CNN's first layer's input channel, resulting in an input tensor of size $x \in \mathbb{R}^{B\times H\times W\times N_S}$ where $B$ is the batch size and $H$ and $W$ are spatial dimensions. This has been proposed for \textit{ex-vivo} head and neck cancer classification \cite{halicek2017deep}. We refer to this method as \textbf{2D CNN HSI}. A disadvantage of this approach is that the entire spectral information vanishes after the first layer. The CNN computes multiple weighted averages of the spectral bands which are simply treated as the CNN's feature dimension afterward. 

This motivates the use of explicit processing methods of the spectral dimension. One approach is to treat the spectral dimension equivalently to a spatial dimension and convolving over it, as employed by \cite{halicek2018optical}. Thus, the CNN's input tensor changes from $x \in \mathbb{R}^{B\times H\times W\times N_S}$ to $\hat{x} \in \mathbb{R}^{B\times H\times W\times N_S \times 1}$ with a channel dimension of one. Here, the 2D CNN architecture can be directly extended to 3D by employing 3D instead of 2D convolutions. We refer to this method as \textbf{3D CNN HSI}.

An approach that has not been considered for head and neck cancer detection is the use of recurrent processing for the spectral dimension, adopted from sequential, temporal learning problems. One approach is to use a convolutional LSTM that performs joint spatial and spectral processing in its units, where each spectral image is processed sequentially. Liu et al. employed this approach in the remote sensing area using a bi-directional model \cite{liu2017bidirectional}. We adopt a similar method using the more efficient gated recurrent units \cite{cho2014learning}, also employing convolutions within its gates. We refer to this method as \textbf{CGRU HSI}.

\begin{figure}[t]
\centering
\includegraphics[width=1.0\textwidth]{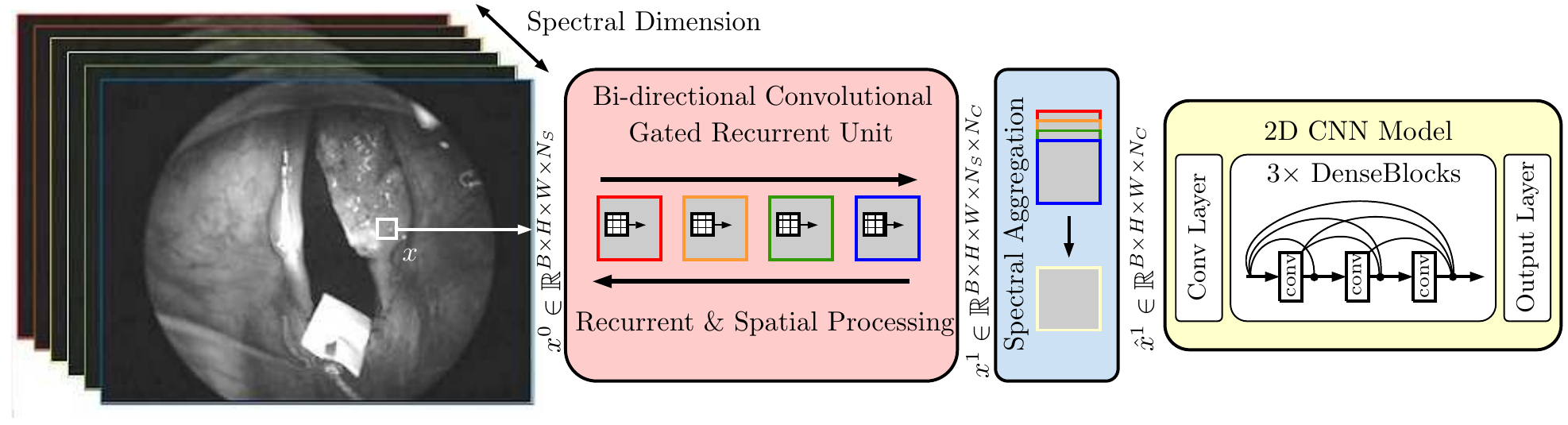}
\caption{The CGRU-CNN we propose for hyperspectral image data. The model's output is a binary classification into benign and malignant.}
\label{fig:model}
\end{figure}

While the CGRU HSI model performs some spatial processing with convolutional gates, high-level, abstract representations are difficult to learn due to limited network depth \cite{liu2017bidirectional}. Therefore, we extend this model by using a 2D CNN for spatial processing after a CGRU layer. Thus, the CGRU receives $x^0 \in \mathbb{R}^{B\times H\times W\times N_S}$ as its input and outputs $x^1 \in \mathbb{R}^{B\times H\times W\times N_S \times N_C}$ where $N_C$ is the CGRU's hidden dimension. Then, we employ a pooling or state selection function $\hat{x}^1 = f_{\mathit{sel}}(x^1)$ that aggregates the tensor's $N_S$ processed spectral states into a single representation $\hat{x}^1 \in \mathbb{R}^{B\times H\times W\times N_C}$. We either select the last spectral state as the output or perform mean or max pooling over all states. Using the last state is a common approach as it captures condensed information from all states, however, state pooling might provide additional information by explicitly combining all states \cite{chen2018enhancing}. The resulting representation is then processed by a 2D CNN with the same structure as 2D CNN RGB or 2D CNN HSI. The architecture \textbf{CGRU-CNN} is depicted in Figure~\ref{fig:model}. 

All our CNN-based models follow the concept of densely-connected convolutional networks \cite{Huang2017}. After an initial convolutional layer, three DenseBlocks follow. Within each DenseBlock, several convolutional layers are employed which make efficient reuse of already computed features by processing all outputs from preceding layers. In between DenseBlocks, an average pooling layer performs spatial and, in case of a 3D CNN, spectral downsampling with a stride of two. After the last DenseBlock, a global average pooling layer is employed, followed by a fully-connected layer that performs binary classification. We train all models with a cross-entropy loss and Adam optimizer. We perform class balancing by reweighting the loss function with the inverse class frequency $w_i = \frac{N}{N_i}$ where $N$ is the total number of samples and $N_i$ the number of samples for class $i$. Hyperparameters such as feature map size and learning rate are optimized for the individual models based on validation performance to provide a fair comparison between all methods. Similarly, we tune RF and SVM hyperparameters including maximum tree depth, the number of trees (RF), the box constraint and kernel (SVM).

We consider the F1-Score, sensitivity, specificity and the area under the receiver operating curve (AUC) as our performance metrics. We do not consider traditional accuracy as it highly misleading for an extremely imbalanced problem such as ours. We report $\SI{95}{\percent}$ confidence intervals (CI) using bias corrected and accelerated bootstrapping with $n_{\mathit{CI}} = \num{10000}$ bootstrap samples. We test for statistically significant difference in our performance metrics using a permutation test with $n_{\mathit{P}} = \num{10000}$ samples and a significance level of $\alpha = 0.05$ \cite{efron1994introduction}.

\section{Results}
\begin{figure}[t]
\centering
\includegraphics[width=0.55\textwidth]{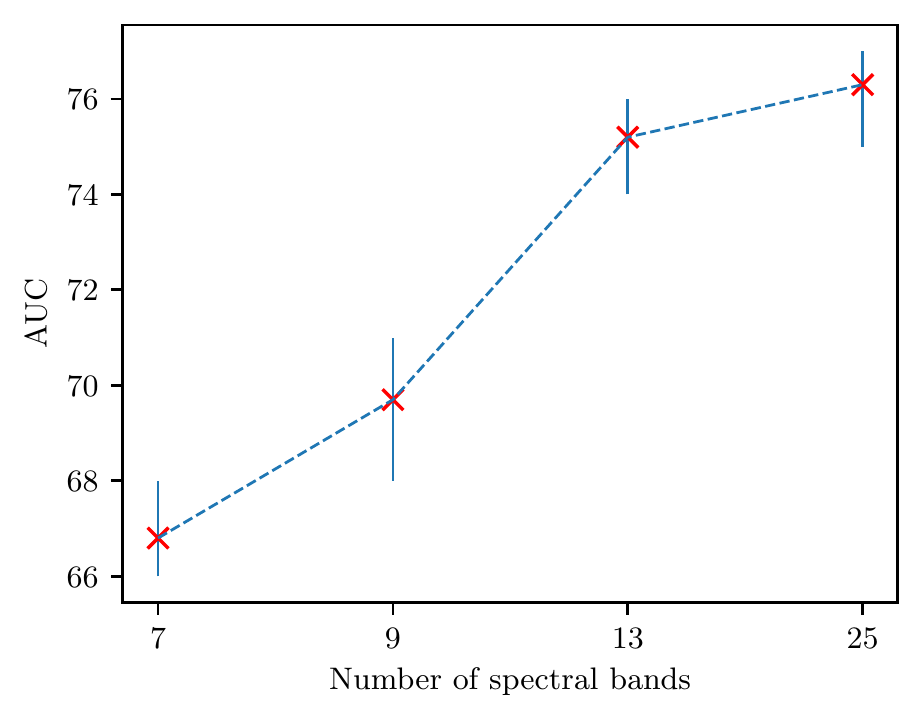}
\caption{Comparison of the AUC with CIs for models trained with a different number of spectral bands with our CGRU-CNN (Last) architecture. For reduction, we selected every second, third and fourth spectral band.}
\label{fig:spectral_dims}
\end{figure}

\begin{table}[t]
\setlength{\tabcolsep}{5pt}
 {\caption{Results for all experiments, comparing different methods. All values are given in percent. $\SI{95}{\percent}$ CIs are provided in brackets. Last, Mean and Max refer to the spectral aggregation strategy.} \label{tab:All-networks-with metrics}}%
\centering
  {\begin{tabular}{l l l l l}
  & \bfseries AUC & \bfseries Sensitivity & \bfseries Specificity & \bfseries F1-Score  \\ \hline
SVM HSI & $60.5(59-61)$ & $58.9(58-60)$ &  $61.9(61-63)$ & $60.8(60-61)$  \\   
RF HSI \cite{laffers2016early} & $59.6(59-60)$ & $71.7(70-73)$ &  $47.5(46-49)$ & $62.7(62-64)$  \\  
2D CNN RGB & $61.2(60-62)$ & $68.5(67-70)$ &  $52.2(51-54)$ & $58.9(58-60)$  \\
2D CNN HSI \cite{halicek2017deep} & $62.3(61-64)$ & $60.0(58-61)$ & $55.1(54-56)$ & $57.6(57-58)$  \\
CGRU HSI \cite{liu2017bidirectional} & $63.5(62-65)$ & $61.0(59-63)$ &  $61.0(59-62)$ & $61.4(60-62)$  \\
3D CNN HSI \cite{halicek2018optical} & $67.8(67-69)$ & $65.7(64-67)$ &  $59.6(58-61)$ & $62.5(62-64)$ \\
CNN-CGRU & $69.9(69-71)$ & $72.4(71-74)$ &  $61.9(61-63)$ & $66.4(65-67)$  \\
CGRU-CNN (Last) & $\pmb{76.3(75-77)}$ & $69.4(68-71)$ &  $\pmb{68.9(68-70)}$ & $69.5(69-71)$  \\
CGRU-CNN (Mean) & $73.7(73-75)$ & $\pmb{74.5(73-76)}$ &  $63.2(62-64)$ & $67.8(67-69)$  \\
CGRU-CNN (Max) & $70.9(70-72)$ & $74.3(73-77)$ &  $66.9(65-68)$ & $\pmb{70.1(69-71)}$  \\
\hline
  \end{tabular}}
\end{table}
The results of our experiments are shown in Table~\ref{tab:All-networks-with metrics}. Using RGB and HSI in a conventional 2D CNN leads to similar performance. Notably, RF and SVM achieve performance in a similar range. When using spatio-spectral processing with a 3D CNN performance improves significantly for the AUC. Overall, variants of our proposed CGRU-CNN architecture performs best. The difference in the AUC, F1-Score and specificity is significant ($p < 0.05$) compared to the previously proposed model 3D CNN. Comparing spectral aggregation strategies, performance varies for different metrics but remains consistently higher than for the other methods. Performing spatial processing first with CNN-CGRU performs better than 3D CNN while being significantly outperformed by CGRU-CNN in terms of the AUC and F1-Score ($p < 0.05$). In Figure~\ref{fig:spectral_dims} we show the effect of using a different number of spectral dimensions with our CGRU-CNN model. When using fewer spectral dimensions the AUC is slightly reduced. For half the number of spectral bands, performance remains similar to using all all spectral bands.

\section{Discussion}

We address the problem of \textit{in-vivo} tumor type classification with hyperspectral imaging. Previously, this task has been addressed for \textit{ex-vivo} data with deep learning methods have shown promising results \cite{halicek2018optical}. However, the \textit{in-vivo} setting makes the task much more difficult due to high variability in terms of motion, lighting and surrounding anatomy. In terms of methods, 2D CNNs and spatio-spectral 3D CNNs have been used for tissue classification. Yet, the individual importance of spatial and spectral information as well as machine learning model's ability to use the information is still unclear, especially under challenging \textit{in-vivo} conditions. Therefore, we employ several machine learning methods to study the information content of spatial and spectral dimensions and whether models can effectively utilize the information. 

Considering our results in Table~\ref{tab:All-networks-with metrics}, we observe a similar performance of 2D CNNs and conventional methods. First, the surprisingly good performance of RFs and SVMs indicates that conventional methods are effectively using spectral information, while not taking spatial relations into account. Second, 2D CNN RGB demonstrates that using spatial information only can also lead to a similar performance level. As a result, we observe that both spectral information alone (RF/SVM) and spatial information with few bands (2D CNN RGB) are valuable for the task at hand, suggesting that combining both is reasonable. 

Using 2D CNNs with HSI in the color channels does not improve performance, indicating that other processing techniques are required. The use of CGRU only, as previously performed for remote sensing \cite{liu2017bidirectional}, does not lead to major improvements either. While simultaneously processing spatial information with convolutions and spectral information with a recurrent gating mechanism, the architecture is very shallow and therefore likely limited in terms of its capability to learn expressive spatial features. This motivates the use of 3D CNNs that treat both spatial and spectral information equally, processing both with convolutions. We observe a significant improvement in the AUC for this approach over the previously mentioned methods. However, treating spatial and spectral information completely equally is also problematic as spectral bands are highly correlated and very similar. Therefore, we propose to use a CGRU-CNN architecture that disentangles spectral and spatial processing by first learning a spatial representation from spectral bands that is then processed by a 2D CNN. This approach significantly outperforms all other methods in terms of the AUC and F1-Score. Interestingly, performing spectral processing after spatial processing with CNN-CGRU performs worse. This suggests that reducing spectral redundancy first by learning a compact representation is preferable. Comparing spectral aggregation strategies, all variations perform consistently well. Using the last state performs best for the AUC, suggesting that the CGRU's learned representation effectively captures relevant information and pooling all spectral states is not required.

Last, we investigate the value of the spectral dimension in more detail. We pose the hypothesis, that spectral bands, while adding valuable information, might be redundant due to high similarity and correlation. While learning a compact spectral representation is one approach, directly removing spectral bands should also reduce redundancy in this regard. The results in Figure~\ref{fig:spectral_dims} show that taking only half the spectral bands still performs very well, indicating that there is indeed redundancy. When reducing the spectral dimension further, however, performance drops notably. This signifies once again that additional spectral information is helpful but too much redundancy limits performance improvement.

Overall, an AUC of $\SI{76.3}{\percent}$ demonstrates the feasibility of \textit{in-vivo} tumor type classification but is likely too limited for clinical decision support yet. Still, we find valuable insights on different machine learning models' capability of utilizing spatial and spectral information. The concept of recurrent-convolutional models has been used for temporal data \cite{gessert2019spatio} which we adopt and extend by bi-directional units and state aggregation methods. Thus, we design a model that demonstrates the concept's efficacy for a hyperspectral learning problem which could substantially improve other medical applications of hyperspectral imaging. 

\section{Conclusion}
We demonstrate the feasibility of \textit{in-vivo} tumor type classification using hyperspectral imaging with spectral and spatial deep learning methods. We design and compare several different methods for processing hyperspectral images to study the value of spatial and spectral information in a challenging \textit{in-vivo} context. Our analysis indicates that processing the spectral dimension with a convolutional gated recurrent network followed by spatial processing with a CNN is preferable. Particularly, conventional and previous deep learning methods are clearly outperformed. These results should be considered when using HSI for tissue classification in future work.

\textbf{Acknowledgments.} This work was partially supported by Forschungszentrum Medizintechnik Hamburg (Grant: 02fmthh2019).

%
%
\bibliographystyle{splncs04}
\bibliography{report}

\end{document}